\documentclass[12pt]{article}
\usepackage{epsfig}
\usepackage{axodraw}
\usepackage{amsmath}
\newcommand{\as}{\alpha_{\mathrm{s}}}
\newcommand{\aem}{\alpha_{\mathrm{em}}}
\renewcommand{\d}{\mathrm{d}}
\renewcommand{\u}{\mathrm{u}}

\renewcommand{\c}{\mathrm{c}}

\newcommand{\X}{\mathbf{X}}
\newcommand{\J}{\mathrm{J}}
\newcommand{\e}{\mathrm{e}}

\newcommand{\g}{\mathrm{g}}
\newcommand{\p}{\mathrm{p}}
\newcommand{\q}{\mathrm{q}}

\newcommand{\qbar}{\mathrm{\overline{q}}}
\newcommand{\ubar}{\mathrm{\overline{u}}}

\newcommand{\kT}{k_{\perp}}
\newcommand{\pT}{p_{\perp}}

\newcommand{\gast}{\gamma^*}
\newcommand{\ga}{\gamma}

\newcommand{\mr}{\mathrm}

\newcommand{\ra}{\rightarrow}
\newcommand{\gtrsim}{\raisebox{-0.8mm}%
{\hspace{1mm}$\stackrel{>}{\sim}$\hspace{1mm}}}

\def\Journal#1#2#3#4{{#1}{\bf #2} (#3) #4}

\def\NPB{{\it Nucl. Phys.~}{\bf B}}
\def\PLB{{\it Phys. Lett.~}{\bf  B}}
\def\JournalPLB#1#2#3{{\it Phys. Lett.~}{\bf {#1}B} (#2) #3}

\def\PRD{{\it Phys. Rev.~}{\bf D}}
\def\ZPC{{\it Z. Phys.~}{\bf C}}

\def\CPC{\it Computer Phys. Commun.~}
\def\PRP{\it Phys. Rept.~}
\def\PRV{\it Phys. Rev.~}
\def\EPJC{{\it Eur. Phys. J.~}{\bf C}}

\def\SJNP{\it Sov. J. Nucl. Phys.~}
\def\SPJP{\it Sov. Phys. JETP~}

{\end{list}}
\newcounter{enumct}

\newlength{\abstwidth}
\newlength{\captivewidth}
\newcommand{\captive}[1]{\rule{5mm}{0mm}%
\begin{minipage}{\captivewidth}%
\caption[small]{#1}\end{minipage}}
\begin{document}
%
 
\sloppy
 
\pagestyle{empty}
 
\begin{flushright}
LU TP 00--31\\
hep-ph/0009003\\
August 2000
\end{flushright}
 
\vspace{\fill}
 
\begin{center}
{\LARGE\bf Effects of Longitudinal Photons}\\[10mm]
{\Large 
Christer Friberg$^\star$ 
and Torbj\"orn Sj\"ostrand$^\star$
}\\[2mm]
{\it Department of Theoretical Physics,}\\[1mm]
{\it Lund University, Lund, Sweden}
\end{center}
 
\vspace{\fill}
\begin{center}
{\bf Abstract}\\[2ex]
\begin{minipage}{\abstwidth}
The description of longitudinal photons is far from trivial,
and their phenomenological importance is largely unknown.
While the cross section for direct interactions is calculable,
an even more important contribution could come from resolved
states. In the development of our model for the interactions of
(real and) virtual photons, we have modeled resolved longitudinal 
effects by simple multiplicative factors on the resolved 
transverse-photon contributions. Recently, a first set of parton 
distributions for longitudinal virtual photons has been presented 
by Ch\'yla. We therefore compare their impact on some representative 
distributions, relative to the simpler approaches.
\end{minipage}
\end{center}

\vspace{\fill}

\footnoterule
{\footnotesize $^\star$christer@thep.lu.se, torbjorn@thep.lu.se}

\clearpage
\pagestyle{plain}
\setcounter{page}{1}

\section{Introduction}

The interactions of a real photon are nontrivial, since the photon 
can fluctuate into partly non-perturbative $\q\qbar$  hadron-like 
states. These strongly-interacting states actually are responsible 
for the bulk of the $\gamma\p$ and $\gamma\gamma$ cross sections.
In the spirit of hadronic physics, it is necessary to introduce 
parton distribution functions (PDF's) in order to describe jet 
production, with non-perturbative boundary conditions at some low 
$\mu_0^2$ scale followed by a perturbatively defined evolution towards 
larger $\mu^2$. Also like in hadronic physics, a description of total, 
elastic and diffractive cross sections tends to rely on Regge 
type phenomenology, but again with many unknowns. The hadron-like 
interaction types are complemented by the direct interactions of the
photons, to form a photoproduction framework.

In moving from the real to the virtual photon, matters do not become
any simpler. Granted, once the photon virtuality $Q^2$ is very
large, the deeply inelastic scattering (DIS) language of $\e\p$
collisions can be used quite successfully, but this language bears
little resemblance with the one used for real photons, and cannot be 
extended to small $Q^2$. So, in the intermediate region, say 
$Q^2 \sim m_{\rho}^2$, physics is at best a bit of each, at worst
beyond either of the two frameworks. In a set of two recent articles 
\cite{ourjet,ourtot} we have tried to develop a model that should 
provide a smooth interpolation between the photoproduction and 
the DIS regions, in the first article studying jet production and in
the second the total cross section of events.

A special problem here is the contribution from longitudinal
photons. By gauge invariance we know that longitudinal photon interactions
must vanish in the limit $Q^2 \to 0$. However, in the few instances
where their effects have been measured at nonvanishing $Q^2$, the 
contribution has been quite significant, from vector meson polarization
in exclusive reactions \cite{longgastpol} to the standard DIS analysis 
of $R = \sigma_{\mathrm{L}}^{\gast\p} / \sigma_{\mathrm{T}}^{\gast\p}$ 
at small $x$ and $Q^2$ \cite{longgastR}. Furthermore, in QED processes
like $\e^+\e^- \to \e^+\e^-\e^+\e^-$ the longitudinal contributions
are 
non-negligible \cite{ChylaFl,siggaga,longgastQED}. Some 
of these contributions can be given a partonic interpretation, so it 
would be tempting again to use the time-honoured parton-distribution
language. This is not a universally accepted route, but we may recall
the successes in describing much of the diffractive phenomenology in
terms of a partonic structure of the Pomeron \cite{IngSch}, a state about 
as virtual and elusive as the longitudinal photon. 

Recently a first set of PDF's for the longitudinal photon was presented 
by Ch\'yla \cite{ChylaFl}. While the analysis is sensitive to the
non-perturbative input for predictions at small $Q^2$, this sensitivity is 
reduced for larger $Q^2$. Access to sensible PDF's should allow an improved 
predictivity for a number of observables, compared with our previous 
approach of using simple $Q^2$- and $\mu^2$-dependent factors to estimate 
the potential impact of the longitudinal-photon contribution. It is 
noteworthy that the total $\gast\p$ and $\gast\gast$ cross sections do 
contain low-$\pT$ components as well, for which a partonic interpretation 
is only implicit, and where the need for simpler ans\"atze remains. 

The plan of this letter is the following. In section 2 we summarize the 
main features of our model for virtual-photon interactions, and in section 3
how the model can be extended to encompass an assumed longitudinal-photon
contribution as well. Some comparisons, between the more sophisticated
approach of having PDF's for longitudinal photons and the simpler one
of $x$-independent multiplicative factors, are presented in section 4. 
Finally, some conclusions are drawn in section 5. 

\section{A Model for Photon Interactions}

In this section we summarize the model presented in \cite{ourjet,ourtot}.
It starts from the model for real photons in \cite{SaSmodel}, but further
develops this model and extends it also to encompass the physics of
virtual photons. The physics has been implemented in the \textsc{Pythia}
generator \cite{pythia}, so that complete events can be studied under 
realistic conditions.   

Photon interactions are complicated since the photon wave function
contains so many components, each with its own interactions. To
first approximation, it may be subdivided into a direct and a resolved
part. (In higher orders, the two parts can mix, so one has 
to provide sensible physical separations between the two.)
In the former the photon acts as a pointlike particle, 
while in the latter it fluctuates into hadronic states.
These fluctuations are of $\mathcal{O}(\alpha_{\mathrm{em}})$, and so
correspond to a small fraction of the photon wave function, but this
is compensated by the bigger cross sections allowed in strong-interaction
processes. For real photons therefore the resolved processes dominate
the total cross section, while the pointlike ones take over for 
virtual photons. 
 
The fluctuations $\gamma \to \q\qbar \, (\to \gamma)$ can be characterized 
by the transverse momentum $\kT$ of the quarks, or alternatively by some
mass scale $m \simeq 2 \kT$, with a spectrum of fluctuations 
$\propto \d\kT^2/\kT^2$. The low-$\kT$ part cannot be calculated 
perturbatively, but is instead parameterized by experimentally determined  
couplings to the lowest-lying vector mesons, $V = \rho^0$, $\omega^0$, 
$\phi^0$ and $\J/\psi$, an ansatz called VMD for Vector Meson 
Dominance. Parton distributions are defined with a unit
momentum sum rule within a fluctuation \cite{SaSpdf}, giving rise
to total hadronic cross sections, jet activity, multiple interactions 
and beam remnants as in hadronic interactions. In interactions with a hadron 
or another resolved photon, jet production occurs by typical 
parton-scattering processes such as $\q\q' \to \q\q'$ or $\g\g \to \g\g$.

States at larger $\kT$
are called GVMD or Generalized VMD, and their contributions to the 
parton distribution of the photon are called anomalous. Given a dividing 
line $k_0 \simeq 0.5$~GeV to VMD states, the anomalous parton distributions 
are perturbatively calculable. The total cross section of a state is not, 
however, since this involves aspects of soft physics and eikonalization 
of jet rates. Therefore an ansatz is chosen where the total cross section 
of a state scales like $k_V^2/\kT^2$, where the adjustable parameter 
$k_V \approx m_{\rho}/2$ for light quarks. The spectrum of GVMD states is taken 
to extend over a range $k_0 < \kT < k_1$, where $k_1$ is identified with 
the $p_{\perp\mathrm{min}}(s)$ cut-off of the perturbative jet spectrum in 
hadronic interactions, $p_{\perp\mathrm{min}}(s) \approx 1.5$~GeV at typical 
energies \cite{pythia}. Above that range, the states are assumed to be 
sufficiently weakly interacting that no eikonalization procedure is required,
so that cross sections can be calculated perturbatively without any recourse
to Pomeron phenomenology. There is some arbitrariness in that choice, and 
some simplifications are required in order to obtain a manageable description.
 
A real direct photon in a $\gamma\p$ collision can interact with the parton
content of the proton: $\gamma\q \to \q\g$ (QCD Compton) and 
$\gamma\g \to \q\qbar$ (Boson Gluon Fusion). The $\pT$ in this collision is 
taken to exceed $k_1$, in order to avoid double-counting with the interactions 
of the GVMD states. In $\gamma\gamma$, the equivalent situation is called 
single-resolved, where a direct photon interacts with the partonic component
of the other, resolved photon. The $\gamma\gamma$ direct process
$\gamma\gamma \to \q\qbar$ has no correspondence in $\gamma\p$.

\begin{figure}[t]
\begin{center}
\begin{picture}(105,200)(-4,-25)
  \Photon(7,140)(45,120){4}{4}  
  \GOval(12,10)(10,5)(0){0.5}
  \DashLine(17,5)(90,5){4}
  \ArrowLine(17,13)(45,30)
  \Gluon(45,75)(45,30){4}{4}
  \ArrowLine(45,75)(45,120)
  \ArrowLine(45,30)(90,30)
  \ArrowLine(90,75)(45,75)
  \ArrowLine(45,120)(90,120) 
  \Text(0,10)[]{$\p$}
  \Text(0,140)[]{$\ga$}     
  \Text(60,50)[]{$\pT$} 
  \Text(60,95)[]{$\kT$} 
  \Text(97,30)[]{$\q'$} 
  \Text(97,75)[]{$\qbar$} 
  \Text(97,120)[]{$\q$} 
\end{picture}  
\hspace{2cm}
\begin{picture}(200,200)(0,0)
  \LongArrow(15,15)(185,15)
  \Text(195,15)[]{$\kT$} 
  \LongArrow(15,15)(15,185) 
  \Text(15,194)[]{$\pT$} 
  \SetWidth{1.5}
  \Line(50,15)(50,180)
  \Text(50,5)[]{$k_0$}
  \Line(100,15)(100,100)
  \Text(100,5)[]{$k_1$}
  \Line(100,100)(180,180)
  \Text(190,190)[]{$\kT = \pT$}
  \Text(32,90)[]{VMD}
  \Text(75,100)[]{GVMD}
  \Text(140,60)[]{direct}
\end{picture}   
\end{center}
\vspace{2mm}
\captive%
{(a) Schematic graph for a hard $\ga\p$ process, illustrating
the concept of two different scales. 
(b) The allowed phase space for this process, with one subdivision
into event classes.
\label{fig:gammapplane}}
\end{figure}

As an illustration of this scenario, the phase space of $\gamma\p$ events is 
shown in Fig.~\ref{fig:gammapplane}. (A corresponding plot can be made for 
$\gamma\gamma$, but then requires three dimensions.)
Two transverse momentum scales are introduced, namely the 
photon resolution scale $\kT$ and the hard interaction scale $\pT$.
Here $\kT$ is a measure of the virtuality of a fluctuation of the photon 
and $\pT$ corresponds to the most virtual rung of the ladder, 
possibly apart from $\kT$. 
As we have discussed above, the low-$\kT$ region corresponds to
VMD and GVMD states that encompasses both perturbative high-$\pT$ and
non-perturbative low-$\pT$ interactions. Above $k_1$, the region is split 
along the line $\kT = \pT$. When $\pT > \kT$ the photon is resolved by
the hard interaction, as described by the anomalous part of the photon 
distribution function. This is as in the GVMD sector, except that we should 
(probably) not worry about multiple parton--parton interactions. In the
complementary region $\kT > \pT$, the $\pT$ scale is just part of the 
traditional evolution of the proton PDF's up to the scale of $\kT$, and thus
there is no need to introduce an internal structure of the photon. 
One could imagine the direct class of events as extending below $k_1$
and there being the low-$\pT$ part of the GVMD class, only appearing 
when a hard interaction at a larger $\pT$ scale would not preempt it.  
This possibility is implicit in the standard cross section framework.  

If the photon is virtual, it has a reduced probability to fluctuate into 
a vector meson state, and this state has a reduced interaction probability.
This can be modeled by a traditional dipole factor
$(m_V^2/(m_V^2 + Q^2))^2$ for a photon of virtuality $Q^2$, where 
$m_V \to 2 \kT$ for a GVMD state. Putting it all together, the cross
section of the GVMD sector then scales like
\begin{equation}
\int_{k_0^2}^{k_1^2} \frac{\d\kT^2}{\kT^2} \, \frac{k_V^2}{\kT^2} \,
\left( \frac{4\kT^2}{4\kT^2 + Q^2} \right)^2 ~.
\end{equation}

For a virtual photon the DIS process $\gast \q \to \q$
is also possible, but by gauge invariance its cross section must
vanish in the limit $Q^2 \to 0$. At large $Q^2$, the direct processes 
can be considered as the $\mathcal{O}(\as)$ correction to the lowest-order 
DIS process, but the direct ones survive for $Q^2 \to 0$. There is no 
unique prescription for a proper combination at all $Q^2$, but we have
attempted an approach that gives the proper limits and minimizes 
doublecounting. For large $Q^2$, the DIS $\gast\p$ cross section
is proportional to the structure function $F_2 (x, Q^2)$ with the
Bjorken $x = Q^2/(Q^2 + W^2)$. Since normal parton distribution 
parameterizations are frozen below some $Q_0$ scale and therefore do not
obey the gauge invariance condition, an ad hoc factor 
$(Q^2/(Q^2 + m_{\rho}^2))^2$ is introduced for the conversion from 
the parameterized $F_2(x,Q^2)$ to a $\sigma_{\mathrm{DIS}}^{\gast\p}$:
\begin{equation}
\sigma_{\mathrm{DIS}}^{\gast\p} \simeq  
\left( \frac{Q^2}{Q^2 + m_{\rho}^2} \right)^2 \,
\frac{4\pi^2\aem}{Q^2} F_2(x,Q^2) =
\frac{4\pi^2\aem Q^2}{(Q^2+m_\rho^2)^2} \,
\sum_{\q} e_{\q}^2 \, \left\{ x  q(x, Q^2) + x \overline{q}(x,Q^2) \right\} 
~.
\label{sigDIS}
\end{equation}
Here $m_\rho$ is some non-perturbative hadronic mass parameter, for 
simplicity identified with the $\rho$ mass. One of the $Q^2/(Q^2+m_\rho^2)$
factors is required already to give finite $\sigma_\mr{tot}^{\ga\p}$ for
conventional parton distributions, and could be viewed as a screening 
of the individual partons at small $Q^2$. The second factor is chosen to give
not only a finite but actually a vanishing $\sigma_\mr{DIS}^{\gast\p}$ 
for $Q^2 \ra 0$ in order to retain the pure photoproduction description there.
This latter factor thus is more a matter  of convenience, and other approaches
could have been pursued.

In order to avoid double-counting between DIS and direct events, a requirement 
$\pT > \max(k_1, Q)$ is imposed on direct events. In the remaining DIS ones, 
denoted lowest order (LO) DIS, thus $\pT < Q$. This would suggest a subdivision
$\sigma_{\mr{LO\,DIS}}^{\gast\p} = \sigma_{\mr{DIS}}^{\gast\p} -
\sigma_{\mr{direct}}^{\gast\p}$, with $\sigma_{\mr{DIS}}^{\gast\p}$ given by
eq.~(\ref{sigDIS}) and $\sigma_{\mr{direct}}^{\gast\p}$ by the perturbative
matrix elements. In the limit $Q^2 \to 0$, the DIS cross section is now
constructed to vanish while the direct is not, so this would suggest
$\sigma_{\mr{LO\,DIS}}^{\gast\p} < 0$. However, here we expect the correct 
answer not to be a negative number but an exponentially suppressed one, 
by a Sudakov form factor. This modifies the cross section: 
\begin{equation}
\sigma_{\mr{LO\,DIS}}^{\gast\p} = \sigma_{\mr{DIS}}^{\gast\p} -
\sigma_{\mr{direct}}^{\gast\p}
~~ \longrightarrow ~~ 
\sigma_{\mr{DIS}}^{\gast\p} \; \exp \left( - \frac{%
\sigma_{\mr{direct}}^{\gast\p}}{\sigma_{\mr{DIS}}^{\gast\p}} \right) \;.
\label{eq:LODIS}
\end{equation}
Since we here are in a region where the DIS cross section is no longer the 
dominant one, this change of the total DIS cross section is not essential. 

\begin{figure}[t]
\begin{center}
\begin{picture}(105,220)(-4,-25)
  \ArrowLine(5,165)(45,165)
  \ArrowLine(45,165)(90,180)
  \Photon(45,165)(45,120){4}{4}  
  \GOval(12,10)(10,5)(0){0.5}
  \DashLine(17,5)(90,5){4}
  \ArrowLine(17,13)(45,30)
  \Gluon(45,75)(45,30){4}{4}
  \ArrowLine(45,75)(45,120)
  \ArrowLine(45,30)(90,30)
  \ArrowLine(90,75)(45,75)
  \ArrowLine(45,120)(90,120) 
  \Text(0,10)[]{$\p$}
  \Text(0,165)[]{$\e$}
  \Text(30,140)[]{$\ga^*$}     
  \Text(30,50)[]{$\g$}     
  \Text(60,50)[]{$\pT$} 
  \Text(60,95)[]{$\kT$} 
  \Text(60,140)[]{$Q$} 
  \Text(97,30)[]{$\q'$} 
  \Text(97,75)[]{$\qbar$} 
  \Text(97,120)[]{$\q$} 
  \Text(97,180)[]{$\e'$} 
\end{picture} 
\hspace{2cm} 
\begin{picture}(200,200)(0,0)
  \LongArrow(15,15)(185,15)
  \Text(195,15)[]{$\kT$} 
  \LongArrow(15,15)(15,185) 
  \Text(15,194)[]{$\pT$} 
  \SetWidth{1.5}
  \Line(15,15)(180,180)
  \Line(100,15)(100,180)
  \Text(100,5)[]{$Q$}
  \Text(190,190)[]{$\kT = \pT$}
  \Text(70,35)[]{LO DIS}
  \Text(150,60)[]{direct}
  \Text(55,100)[]{non-DGLAP}
  \Text(135,170)[]{resolved $\gast$}
\end{picture}   
\end{center}
\vspace{2mm}
\captive%
{(a) Schematic graph for a hard $\ga^*\p$ process, illustrating
the concept of three different scales. (b) Event classification
in the large-$Q^2$ limit.
\label{fig:gastpplane}}
\end{figure}

The overall picture, from a DIS perspective, is illustrated in 
Fig.~\ref{fig:gastpplane},
now with three scales to be kept track of. The traditional DIS region 
is the strongly ordered one, $Q^2 \gg \kT^2 \gg \pT^2$, where 
DGLAP-style evolution \cite{DGLAP} is responsible for the event 
structure. As always, ideology wants strong ordering, while 
the actual classification is based on ordinary ordering 
$Q^2 > \kT^2 > \pT^2$. The region $\kT^2 > \max(Q^2,\pT^2)$ is also
DIS, but of the $\mathcal{O}(\as)$ direct kind. The region 
where $\kT$ is the smallest scale corresponds to 
non-ordered emissions, that then go beyond DGLAP validity,
while the region $\pT^2 > \kT^2 > Q^2$ cover the interactions of a 
resolved virtual photon. Comparing Figs.~\ref{fig:gammapplane}b
and \ref{fig:gastpplane}b, we conclude that the whole region
$\pT > \kT$ involves no doublecounting, since we have made no
attempt at a non-DGLAP DIS description but can choose to cover this 
region entirely by the VMD/GVMD descriptions. Actually, it is only 
in the corner $\pT < \kT < \min(k_1, Q)$ that an overlap can occur 
between the resolved 
and the DIS descriptions. Some further considerations show that
usually either of the two is strongly suppressed in this region,
except in the range of intermediate $Q^2$ and rather small $W^2$.
Typically, this is the region where $x \approx Q^2/(Q^2 + W^2)$ is not 
close to zero, and where $F_2$ is dominated by the valence-quark 
contribution. The latter behaves roughly $\propto (1-x)^n$, with an 
$n$ of the order of 3 or 4. Therefore we will introduce a corresponding 
damping factor to the VMD/GVMD terms. 

In total, we have now arrived at our ansatz for all $Q^2$:
\begin{equation}
\sigma_\mr{tot}^{\gast\p} = 
\sigma_{\mr{DIS}}^{\gast\p} \; \exp \left( - 
\frac{\sigma_{\mr{direct}}^{\gast\p}}{\sigma_{\mr{DIS}}^{\gast\p}} \right) +
\sigma_{\mr{direct}}^{\gast\p} +
\left( \frac{W^2}{Q^2 + W^2} \right)^n \left(
\sigma_{\mr{VMD}}^{\gast\p} + 
\sigma_{\mr{GVMD}}^{\gast\p} \right) \;,
\end{equation}
with four main components. Most of these in their turn have
a complicated internal structure, as we have seen. The $\gast\gast$ 
collision between two inequivalent photons contains 13 components: four 
when the VMD and GVMD states interact with each other (double-resolved),
eight with a LO~DIS or direct photon interaction on a VMD or GVMD state on
either side (single-resolved, including the traditional DIS), and one 
where two direct photons interact by the  process $\gast\gast \to \q\qbar$ 
(direct, not to be confused with the direct process of $\gast\p$). 

An important note is that the $Q^2$ dependence of the DIS and direct 
processes is implemented in the matrix element expressions, i.e. in 
processes such as $\gast\gast \to \q\qbar$ or $\gast \q \to \q\g$
the photon virtuality explicitly enters. This is different from 
VMD/GVMD, where dipole factors are used to reduce the assumed flux of 
partons inside a virtual photon relative to those of a real one, but 
the matrix elements themselves contain no dependence on the virtuality
either of the partons or of the photon itself. Typically results are 
obtained with the SaS~1D PDF's for the virtual (transverse) photons
\cite{SaSpdf}, since these are well matched to our framework, e.g.
allowing a separation of the VMD and GVMD/anomalous components. 

\section{The Longitudinal Photon Contribution}

In $\e\p$ interactions, the cross section can be written as \cite{sigmaTL}: 
\begin{equation}
\frac{\d^2\sigma(\e\p \rightarrow \e\X)}{\d y \, \d Q^2}
=  f_{\ga/\e}^{\mr{T}}(y,Q^2) \sigma_\mr{T}(y,Q^2) + 
f_{\ga/\e}^{\mr{L}}(y,Q^2) \sigma_\mr{L}(y,Q^2) ~,
\label{sigmaTL}
\end{equation}
with the fluxes of transverse and longitudinal photons given by 
\begin{eqnarray}
f_{\ga/\e}^{\mr{T}}(y,Q^2) & = & \frac{\aem}{2\pi} 
\left( \frac{1+(1-y)^2}{y} \frac{1}{Q^2}-\frac{2m_{\e}^2y}{Q^4} \right)\;,\\
f_{\ga/\e}^{\mr{L}}(y,Q^2) & = & \frac{\aem}{2\pi} 
\frac{2(1-y)}{y} \frac{1}{Q^2}\;.
\label{LLogflux}
\end{eqnarray}
Here $y=qP/kP$ is the energy-momentum fraction carried off from the incoming 
electron by the virtual photon. The $y$ or $Q^2$ can be traded in for the 
Bjorken $x$, but this $x$ can be given an interpretation in terms of the 
momentum fraction of the struck quark in the proton only in the DIS region 
of large $Q^2$. In $\e^+\e^-$ events, an 
$f_{\ga/\e}^{\mr{T}}$ or $f_{\ga/\e}^{\mr{L}}$ occurs on each side, thus 
giving four terms by simple generalization of eq.~(\ref{sigmaTL}). 

The model summarized in the previous section is intended to describe in detail
the $\sigma_\mr{T}$ term, which is composed of all the many different kinds of
events. So far, nothing has been said about $\sigma_\mr{L}$, except that gauge
invariance dictates its vanishing in the limit $Q^2 \to 0$. However, we will
assume that quite a similar decomposition can be made of longitudinal photon
interactions as was done for the transverse one. To first approximation, 
this again means a separation into direct and resolved photons.
In direct processes, the nature of the photon is explicitly included in the 
perturbative cross section formulae. Thus, for $\gast\q \to \q\g$ and
$\gast \g\to\q\qbar$, the differential cross sections  
$\d\hat{\sigma}_{\mathrm{T}} / \d\hat{t}$ and
$\d\hat{\sigma}_{\mathrm{L}} / \d\hat{t}$ are separately available 
\cite{siggap}. The latter is proportional to $Q^2$ and thus nicely
vanishes in the limit $Q^2 \to 0$. Similarly the $\gast\gast \to \q\qbar$
process gives four separate cross section formulae,
$\d\hat{\sigma}_{\mathrm{TT,TL,LT,LL}} / \d\hat{t}$ \cite{siggaga}.
The DIS, non-direct part currently contains no explicit description of a
longitudinal probing photon, only of a probed one.
However, to the extent that PDF's are extracted from 
$F_2 \propto \sigma_{\mathrm{T}} + \sigma_{\mathrm{L}}$ data,
effects may be implicitly included. Furthermore, perturbative calculations
\cite{RinDISthy} predict $\sigma_{\mathrm{L}} \ll \sigma_{\mathrm{T}}$
in the large-$Q^2$ region, where this process dominates. 

For resolved processes, interactions come in two kinds.\\
\textit{(i)} Given a PDF set for the longitudinal photon, jet cross sections 
can be obtained by the traditional convolution of parton fluxes with the 
hard-scattering matrix elements. The PDF's are to be evaluated at some 
factorization scale
$\mu^2$ related to the hardness of the scattering, e.g. 
$\mu^2 = \pT^2 = \hat{t}\hat{u}/\hat{s}$ if $Q^2$ can be neglected.\\
\textit{(ii)} In low-$\pT$ interactions there is no easily definable 
perturbative scale
$\mu$. The relevant scale instead can be taken as the mass of the state,
i.e. $m_V$ for VMD and $2\kT$ for GVMD, or $m_{\rho}$ if one simplifies 
even further, given that the $\rho^0$ dominates the VMD/GVMD sector (though
less so at large $Q^2$). Such a choice is not unreasonable also from a 
partonic point of view: low-$\pT$ means no interactions above 
$k_1 = p_{\perp\mathrm{min}}(s) \approx 2 m_{\rho}$ but certainly allows
interactions below this scale, and $k_1/2$ might be a reasonable estimate
of the typical order. 

In the past, we have studied a few simple multiplicative expressions. 
Rewriting eq.~(\ref{sigmaTL}) (for the resolved part only) as
\begin{equation}
\frac{\d^2\sigma(\e\p \rightarrow \e\X)}{\d y \, \d Q^2}
=  f_{\ga/\e}^{\mr{T}} \sigma_\mr{T}
\left( 1 + \frac{f_{\ga/\e}^{\mr{L}}}{f_{\ga/\e}^{\mr{T}}}
\, \frac{\sigma_\mr{L}}{\sigma_\mr{T}} \right)
=  f_{\ga/\e}^{\mr{T}} \sigma_\mr{T}
\left( 1 + \frac{f_{\ga/\e}^{\mr{L}}}{f_{\ga/\e}^{\mr{T}}}
\, R \right) ~,
\label{eq:sigT}
\end{equation}
the forms introduced for jet production were
\begin{eqnarray}
R = R_1(y,Q^2,\mu^2) & = & a \frac{4 \mu^2 Q^2}{(\mu^2 + Q^2)^2}
~,\label{Rfact1}\\
R = R_2(y,Q^2,\mu^2) & = & a \frac{4 Q^2}{(\mu^2 + Q^2)}
~,\label{Rfact2}\\
R = R_3(y,Q^2,\mu^2) & = & a \frac{4 Q^2}{(m_{\rho}^2 + Q^2)}
~,\label{Rfact3}
\end{eqnarray}
and for the total cross section processes
\begin{eqnarray}
R = r_1(m_V^2, Q^2) & = & a \frac{4 m_V^2 Q^2}{(m_V^2 + Q^2)^2}
~,\label{rV1}\\
R = r_2(m_V^2, Q^2) & = & a \frac{4 Q^2}{(m_V^2 + Q^2)}
~.\label{rV2}
\end{eqnarray}
Here $a$ in all cases denotes an unknown number, where the intention was
to use $a=1$ as an extreme contrast to the no-longitudinal-effects $a=0$,
with the truth likely to be somewhere in between. All the expressions 
were constructed to vanish like $Q^2$ for $Q^2 \to 0$. $R_1$ and $r_1$
also vanish for large $Q^2$, while the rest there become $Q^2$-independent.
The $\mu$-independent option $R_3$ essentially is the same as $r_2$.
In $\gast\gast$ events, the same approach is pursued, with one 
multiplicative factor for each side with a resolved photon.

It is thus this framework that should be contrasted with what is offered 
by a set $f_i^{\gast_{\mr{L}}}(x, \mu^2)$ of PDF's for the longitudinal 
photon, where the dependence on parton species $i$ and momentum fraction
$x_i$ is explicitly given. Assuming that a hard interaction at scale $\mu^2$ 
is selected for the transverse photon, this entails knowledge of $i$ 
and $x_i$. Then a sensible choice would be
\begin{equation}
R = R_{\mr{PDF}} = \frac{Q^2}{Q^2 + m_{\rho}^2} \, 
\frac{f_i^{\gast_{\mr{L}}}(x_i, \mu^2, Q^2)}%
{f_i^{\gast_{\mr{T}}}(x_i, \mu^2, Q^2)} ~.
\end{equation}
We have here chosen to introduce one modification for the Ch\'yla 
$f_i^{\gast_{\mr{L}}}$, as can be seen. His parameterizations are not intended 
to be valid for $Q^2 < 1$~GeV$^2$ or thereabout, since the quark mass effects 
have not been included, which would provide a dampening in that region. 
In order to use them below that scale, the ad hoc multiplicative factor 
$Q^2/(Q^2 + m_{\rho}^2)$ is introduced to ensure the correct limiting 
behaviour for $Q^2 \to 0$, while rapidly approaching unity for 
$Q^2 > 1$~GeV$^2$. The PDF's can then be frozen below the lowest scale 
for which they can meaningfully be evaluated. 
In the explicit calculation of 
$f_i^{\gast_{\mr{L}}}(x_i, \mu^2, Q^2)$, 
the relations $0.001 \leq x_i \leq 0.995$ and 
$1 \leq \ln(\mu^2/\Lambda^2_\mr{QCD})/\ln(Q^2/\Lambda^2_\mr{QCD}) \leq 3.9$
need to be fulfilled and are set to the relevant boundary values if not.
It may lead to a negative PDF, however, wherefore the requirement 
$f_i^{\gast_{\mr{L}}} \geq 0$ is imposed when calculating $R_\mr{PDF}$.

For the studies in this article, the procedure used is to generate events   
based on the transverse part of resolved photons only, and then to apply
one or several of the $R$ options above as weight for the event. Thus the 
inclusion of resolved longitudinal-photon effects is not seen as the 
appearance of any new kinds of hadronic final states, but only
as a more or less increased cross section for the already existing 
transverse-photon ones. This may not be entirely correct --- any 
initial-state radiation would probe $f_i^{\gast_{\mr{L}}}$ in the
region of scales below $\mu^2$ and momentum fractions above $x_i$, e.g.  
--- but should be a good first approximation, especially in view of all the
other uncertainties. 

\section{Some Results}

The significance of the longitudinal-photon contribution depends on the 
$f_i^{\gast_{\mr{L}}}/f_i^{\gast_{\mr{T}}}$ ratio, but also on other 
factors, such as the size of the direct contribution, or the smearing
of the parton-level kinematics in realistic observables. We therefore
begin by a brief study of $f_i^{\gast_{\mr{L}}}/f_i^{\gast_{\mr{T}}}$
itself before illustrating experimental consequences.

\begin{figure} [t]
   \begin{center}
   \mbox{\psfig{figure=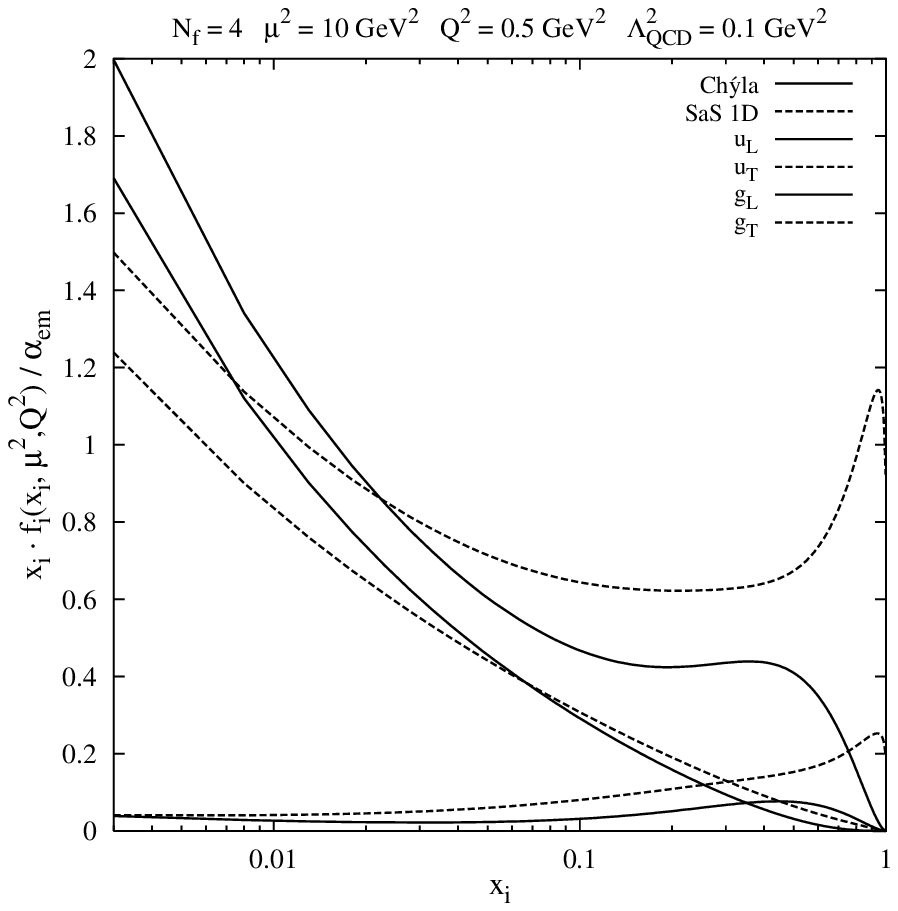,width=105mm}\hspace{-30mm}
	\psfig{figure=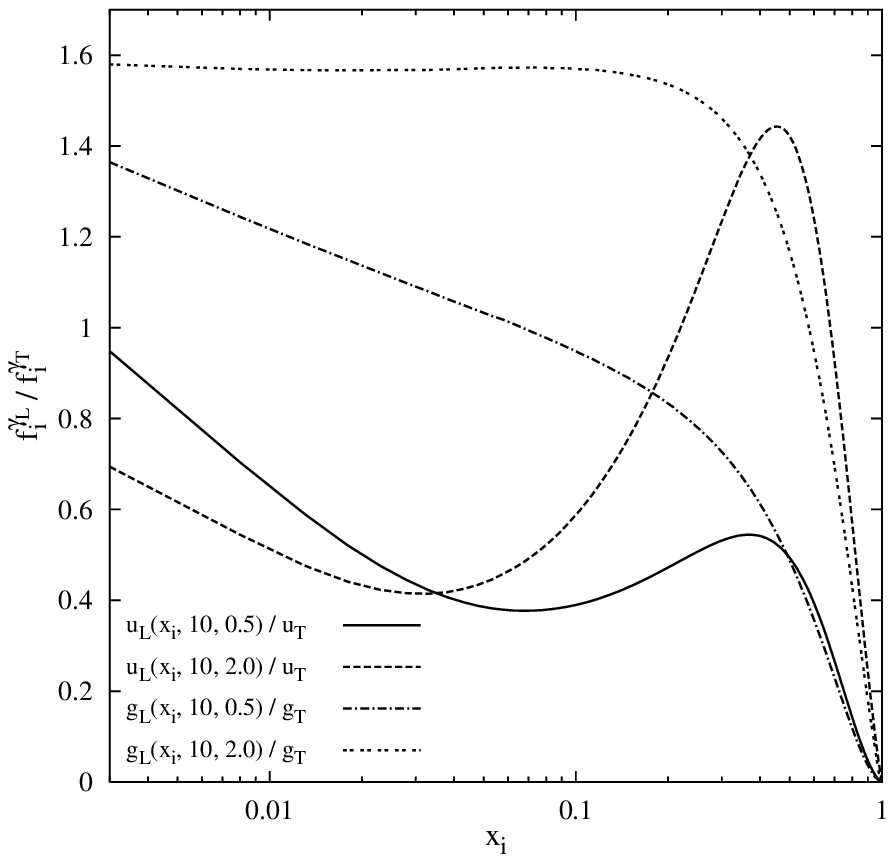,width=105mm}}
   \end{center}
\vspace{-5mm}\hspace{2cm}a)\hspace{8cm}b)\\
\captive{%
a) The $x_i$-weighted parton distribution functions as given by Ch\'yla
and SaS~1D, solid and dashed lines respectively, for a factorization scale 
$\mu^2=10$~GeV$^2$ and a photon virtuality $Q^2=0.5$~GeV$^2$. 
$x_i$ is the momentum fraction carried by parton $i$. In falling order at 
small $x_i$ the following curves are shown: 
the sum of the gluon and four lightest flavours 
(including the respective anti-flavour contribution), 
the gluon and the $\u$-quark (excluding $\ubar$) contributions.  
b) The ratio of the Ch\'yla and SaS~1D PDF's as a function of $x_i$ for the 
$\u$-quark and the gluon at two different photon virtualities, 0.5 and 
2~GeV$^2$, for a fixed factorization scale $\mu^2=10$~GeV$^2$. 
\label{fig:xf}}
\end{figure}

The virtual photon PDF used for transversely resolved photons is the SaS~1D 
one~\cite{SaSpdf}. In Fig.~\ref{fig:xf}a it is compared to Ch\'yla's
longitudinal photon PDF as a function of the momentum fraction $x_i$ carried 
by parton $i$. 
The down-type quarks give one quarter of the $\u$-quark contribution,
due to the difference in electric charge. In Ch\'yla's PDF, no difference 
is made between $\u$ and $\c$, whereas in SaS~1D charm mass effects are 
included, which dampens the distribution at low photon virtualities. 
In general, the low-end region in $x_i$ is dominated by the gluon contribution
and, with the subdivision made in SaS~1D, the major part comes from the VMD 
component (at this low photon virtuality). In the high-$x_i$ end, it is 
instead the valence quarks that dominate and consequently the anomalous, 
point-like, component of the PDF. The longitudinal PDF's increase faster 
as compared to the transverse ones when going to lower $x_i$.

With $\mu^2=10$~GeV$^2$, the $\u$-quark and gluon ratios of 
the two PDF's are shown in Fig.~\ref{fig:xf}b as a function of $x_i$ for two
different photon virtualities, 0.5 and 2~GeV$^2$. At $x_i<0.5$ the ratios 
are between 0.4 and 1.6. When increasing the photon virtuality for a fixed
factorization scale $\mu^2$ a non-trivial change in the ratios is obtained due 
to various effects, for example, shrinking evolution ranges and a faster 
dampening for the VMD component as compared to the anomalous one.

\begin{figure} [t]
   \begin{center}
   \mbox{\psfig{figure=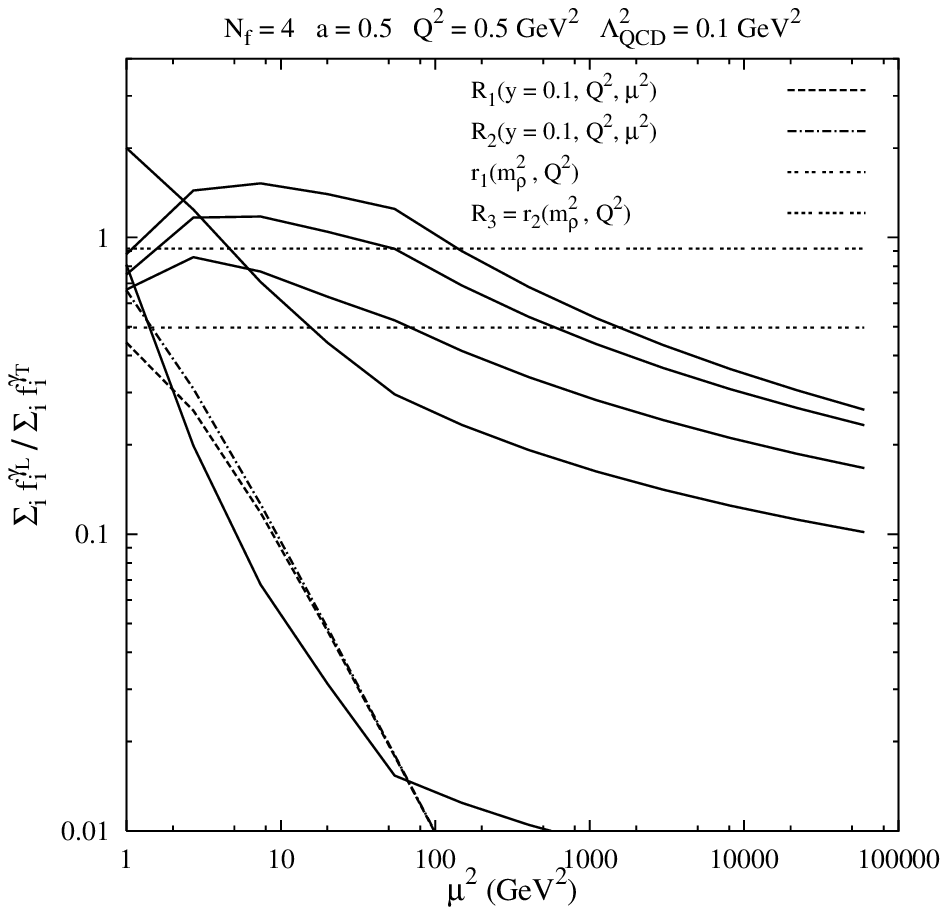,width=105mm}}
   \end{center}
\captive{%
The ratio of longitudinal and transverse parton distributions of the photon,
$\sum_i f_i^{\gast_{\mr{L}}} / \sum_i f_i^{\gast_{\mr{T}}}$, as a function of 
the factorization scale $\mu^2$ for different fixed parton momentum fractions 
$x_i$. The photon virtuality is 0.5 GeV$^2$. 
The ratio is between the sum of the gluon and the four lightest
flavours contribution for the two PDF's.
In decreasing order at high $\mu^2$, $x_i$ is equal to: 
0.001, 0.01, 0.1, 0.5 and 0.9. $a=0.5$ was used to calculate the different 
$R$ factors.
\label{fig:mu}}
\end{figure}

In Fig.~\ref{fig:mu}, the PDF ratio 
$\sum_i f_i^{\gast_{\mr{L}}} / \sum_i f_i^{\gast_{\mr{T}}}$
is plotted as a function of the factorization scale $\mu^2$ for different 
fixed $x_i$ with a photon virtuality of 0.5 GeV$^2$. Also shown are the 
different $R$ factors with $a=0.5$. 
In the complete event generation, the mass of the fluctuation is used in 
the $r_i$ factors (eq.~(\ref{rV1}) and (\ref{rV2})), i.e. $m_V$ for 
VMD and $2\kT$ for GVMD, but for illustrative purposes the $\rho^0$ 
mass has been used for this particular distribution. Therefore, in this
plot, $r_2$ reduces to the simple $R_3$ alternative. 
At small and medium $x$-values the $\mu^2$ dependence is moderate and the
$\mu^2$-independent $r_2$ alternative is a reasonable 
approximation to $R_\mr{PDF}$. With $Q^2=0.5$~GeV$^2$, the $r_1$
alternative is about half of the $r_2$ one. The $\mu^2$-dependent factors,
$R_1$ and $R_2$, do fall off in agreement with the ratio 
$f_i^{\gast_{\mr{L}}} / f_i^{\gast_{\mr{T}}}$ when $x_i=0.9$, but completely
fail at smaller $x_i$. 

We now turn to more realistic distributions, picking one $\gast\p$ and one
$\gast\gast$ example, where additionally the former probes jet
cross sections and the latter total cross sections. The interesting range of
photon virtualities, to study the longitudinal resolved photon effects, 
is at medium $Q^2$, say in the interval $m_{\rho}^2 - 4~\mr{GeV}^2$.  
When approaching $Q^2=0$ the 
longitudinal photon PDF vanish, and at large $Q^2$ the unresolved events,
the direct and DIS ones in our description, take over and dominate the 
cross sections. In the following, the $r_i$ factors will be used according 
to eq.~(\ref{rV1}) and (\ref{rV2}), without the $m_{\rho}$ approximation.

\begin{figure} [t]
   \begin{center}
   \mbox{\psfig{figure=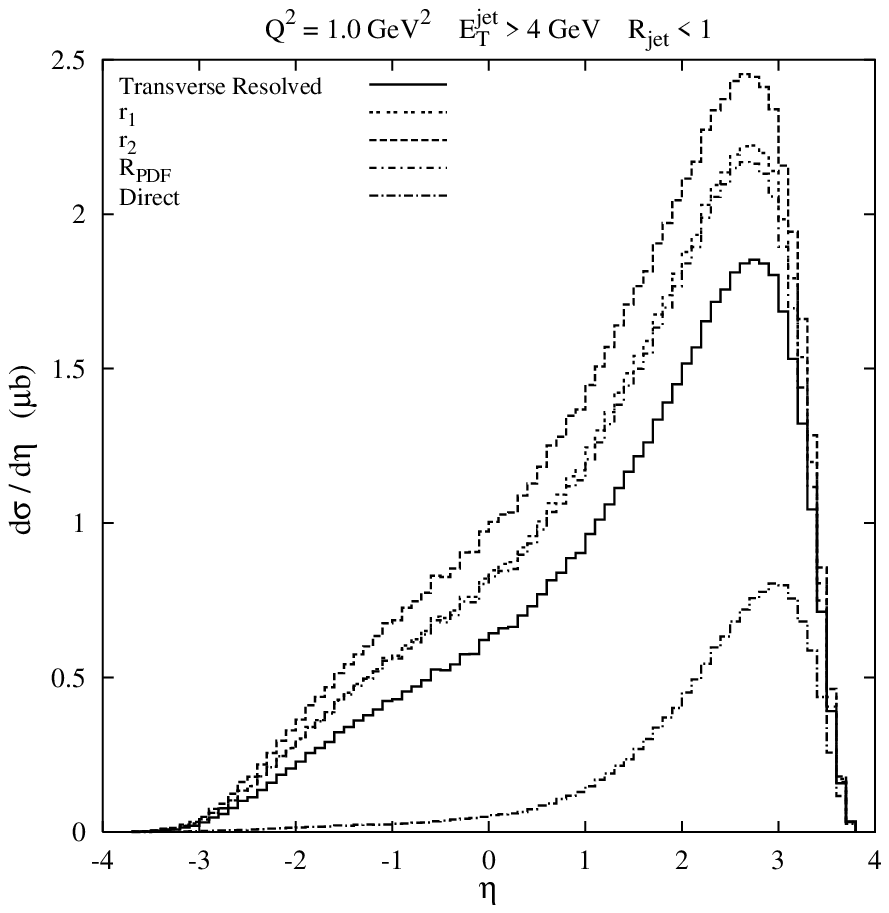,width=99mm}\hspace{-25mm}
	\psfig{figure=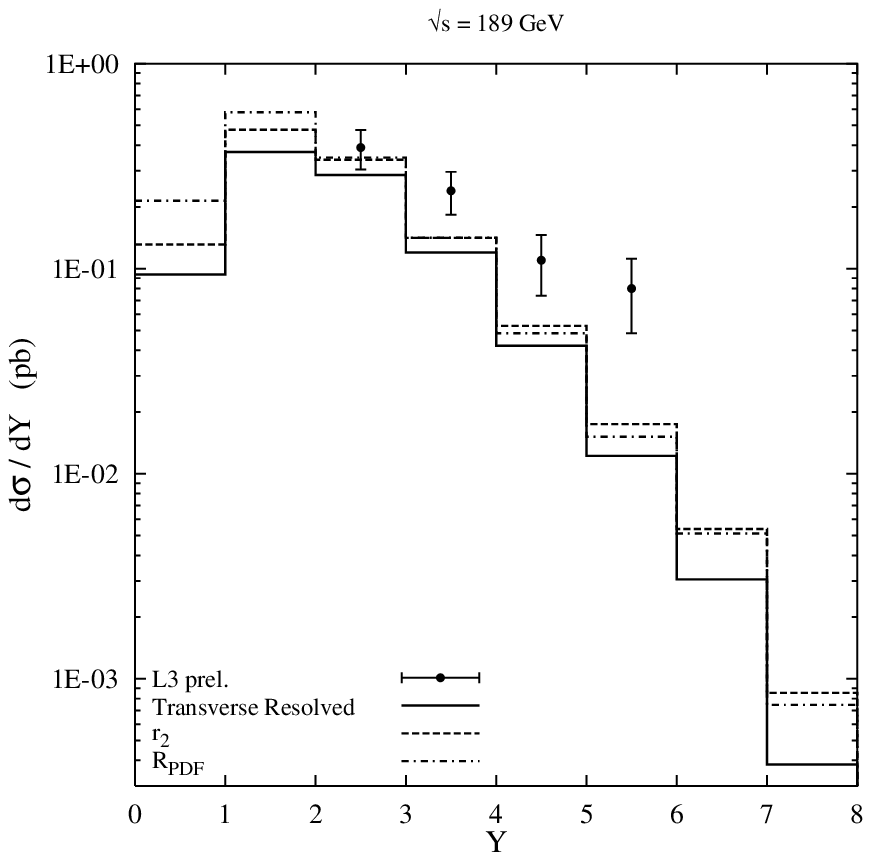,width=105mm}}
   \end{center}
\vspace{-5mm}\hspace{2cm}a)\hspace{8cm}b)\\
\captive{%
a) The $\gast\p$ jet cross section as a function of the pseudo-rapidity $\eta$ 
of the jets. `Direct' shows the
contribution from direct events only. The others show 
the total contribution of all event classes with 
and without $R$ factors to include (or not include) longitudinal resolved 
photon effects. 
$a=0.5$ is used for $r_i$. 
b) The differential $\e^+\e^- \ra \e^+\e^- + \mr{hadrons}$ 
cross section as a function of $Y=\ln (y_1y_2s/\sqrt{Q_1^2Q_2^2})$. Same 
notation as in a).
\label{fig:ee+gp}}
\end{figure}

In $\gast\p$, a cone jet algorithm was used to find jets with
transverse energy $E_{\perp}^\mr{jet} > 4~\mr{GeV}$ 
within a radius $R = \sqrt{(\Delta \eta)^2+(\Delta\phi)^2} < 1$. The
invariant mass of the $\gast\p$ system is $W_{\gast\p}=200~\mr{GeV}$ and
the photon virtuality $Q^2=1~\mr{GeV}^2$. 
At HERA $\sqrt{s_{\e\p}} \simeq 300~\mr{GeV}$, wherefore a fixed $y=0.44$
was used when calculating the ratio between the longitudinal and transverse
photon fluxes $f_{\ga/\e}^\mr{L}(y,Q^2) / f_{\ga/\e}^\mr{T}(y,Q^2)$ in 
eq.~(\ref{eq:sigT}). The proton parton distribution used is the 
CTEQ5L~\cite{CTEQ5L}.

The obtained differential jet cross section with at least one jet, 
with respect to the jet pseudo-rapidity $\eta$, is shown in 
Fig.~\ref{fig:ee+gp}a. The photon direction is along the positive $y$-axes. 
`Transverse Resolved' is the total contribution from the sum of the 
different event classes in our model, but only taking into
account the transversely resolved photon contribution. 
The contribution from direct events, summed over $\sigma_\mr{T,L}$, is 
shown separately as a reference for the part being unaffected by the
different $R$ factors to be applied. The $x_i$-independent $r_1$ factor
and the $R_\mr{PDF}$ factor give about the same enhancement of the jet cross 
section. The jet cross section with the $r_2$ factor is well above the other
two. $a$ was for simplicity chosen to 0.5 when calculating $r_i$. This 
was the value used in ref.~\cite{ourtot} to obtain a nice agreement between 
data and the model for total $\gast\p$ and $\gast\gast$ cross sections with the
$r_1$ alternative, and also there overshot with the $r_2$ one. The requirement
$E_{\perp}^\mr{jet} > 4~\mr{GeV}$, i.e. $\mu^2 \gtrsim 16~\mr{GeV}^2$, 
suppress the $R_1$ and $R_2$ alternatives and give only a small enhancement 
of the jet cross section for them. 

The effects of applying the different $R$ factors above is most pronounced
at central rapidities, where the distribution is not so much contaminated 
by the direct events. With increasing photon virtuality, starting at small
but non-negligible $Q^2$, the jet cross sections obtained with $r_1$ and 
$r_2$ increase faster than the $R_\mr{PDF}$ one 
(all with respect to the `Transverse Resolved' jet cross section). 
At $Q^2=1$~GeV$^2$, as studied above, the $r_1$ 
and $R_\mr{PDF}$ agree, with $r_2$ being above. Continuing to larger 
virtualities, the $R_\mr{PDF}$ jet cross section will continue increasing 
relatively to the `Transverse Resolved', approaching the $r_2$ one,
with the $r_1$ alternative much below. At large photon 
virtualities the effect of the $R$ factors will gradually decrease in 
importance, because the direct cross section will start to dominate 
the event sample. The main point, however, is that a simple 
$x_i$-independent multiplicative factor works well over the whole $Q^2$
range. This holds also in a simultaneous study of two-jet event properties.

The effects of longitudinal resolved photons can be expected to be
important in $\gast\gast$ events, if at least one of the photon virtualities
is not too large. Double-tagged two photon events have been measured by 
the L3 collaboration~\cite{L3} and is shown with respect to the 
variable $Y=\ln(y_1 y_2 s/\sqrt{Q_1^2 Q_2^2})$ in Fig.~\ref{fig:ee+gp}b, 
$y_i$ being the energy-momentum fraction carried by photon $i$, $Q_i^2$
their respective virtuality and $s$ the CM energy squared of the colliding 
$\e^+\e^-$ pair. 

As concluded in ref.~\cite{ourtot}, the direct
contribution is the dominant part of the cross section. 
Again, the `Transverse Resolved' is the total contribution only taking into
account transversely resolved photons. 
The direct events are included in the total contribution with
both transverse and longitudinal cross sections considered. 
The cross section with the $r_2$ alternative for estimating the 
longitudinal resolved photon effects, using $a=0.5$, give about the
same result as with the ratio of the longitudinal and transverse PDF's, 
$R_\mr{PDF}$. The result with $r_1$ (not shown) is between the 
`Transverse Resolved' and $r_2$. 
Elastic, diffractive and low-$p_{\perp}$ events give a negligible 
contribution to this $\gast\gast$ cross section. With the hard scale in the 
processes being relatively large, small factors $R_1$ and $R_2$ are obtained. 
They disagree with the $R_\mr{PDF}$, which is due to the rather small 
typical $x_i$ values in the events (cf. Fig.~\ref{fig:mu}). 
 
At small $Y$, corresponding to medium $x_i$ and large $Q_i^2$, the single- and 
double-resolved events give small contributions, so the major part of the
resolved photon events in this region comes 
from a photon being resolved by a DIS photon, eq.~(\ref{eq:LODIS}) 
(replacing direct with single-resolved). 
In our model, the DIS process is simply the $\gast\q \ra \q$ one and 
therefore only the quark part enters the $R_\mr{PDF}$ factor here. Typical 
$x_i$ values for the lowest $Y$ bins are $\sim 0.5$, corresponding to the 
peak region of Fig.~\ref{fig:xf}b, where 
$f_i^{\gast_{\mr{L}}}/f_i^{\gast_{\mr{T}}}$
is exceptionally large, and this explains why the $r_2$ is much below 
$R_\mr{PDF}$ here. It offers an example of a region where the 
$x_i$-dependence of $f_i^{\gast_{\mr{L}}}/f_i^{\gast_{\mr{T}}}$ cannot 
be ignored. 

At large $Y$, $x_i$ and $Q_i^2$ are smaller (but tagging conditions require
$Q_i^2 > 3$~GeV$^2$), and the single-resolved processes are most important 
among the resolved ones. The low $x_i$ region is gluon-dominated and 
$f_i^{\gast_{\mr{L}}}/f_i^{\gast_{\mr{T}}}$ vary moderately, wherefore a
simple $x_i$-independent factor can be well approximated for the PDF ratio.
While $R_\mr{PDF}$ does increase with $Y$, already the 
transversely resolved photon contributions are too small to give a large 
effect with the $R$ factors included, i.e. the level of the data points
is not reached. 

\section{Summary and Outlook}
The exploration of the photon structure offers challenges, both for 
theory and experiment. The theory challenge is to construct a realistic
scenario, that covers all the aspects that we know or expect to be there.
The experimental challenge is to check each of these aspects separately.

Our photon scenario involves four main components for the real photon ---
direct, DIS, VMD and GVMD, with the latter two corresponding to a discrete
set and a continuum of resolved-photon states, respectively.
For $\gast\gast$ interactions, the number of combinations is thirteen.
Depending on photon virtualities, these are mixed in varying proportions. 
Each state can undergo a set of different interactions, that comprise 
both high-$\pT$ jet production and (for most components) low-$\pT$
physics. The model has several free parameters, such as parton 
distributions, $\mu^2$ scales of hard processes, $\pT$ cut-off scales, 
and non-perturbative primordial $\kT$ distributions.  

Needless to say, the resulting complexity makes experimental tests 
nontrivial. Any single distribution will receive contributions from 
several components, states and interactions, usually with so much overlap 
that it is difficult to distinguish their separate contributions. Some 
distributions may illuminate a special point, like the separation 
between direct and resolved photons in the $x_{\gamma}$ distribution
\cite{xgamma}. Rapidity and $\pT$ distributions of jets, underlying event 
activity, the $\pT$ distribution of the beam remnant jet, and elastic and
diffractive topologies are among other measures offering a partial
separation. In the future, the simultaneous study of many observables
could provide further information, but probably there will be few simple 
answers.

In this article we have studied one further complication, namely the 
poorly-known structure of the resolved longitudinal photon. In principle,
a separation from transverse photons is provided by the difference in $y$ 
dependence between $f_{\ga/\e}^{\mr{L}}$ and $f_{\ga/\e}^{\mr{T}}$,
but few experiments can offer the range of CM energies and tagging 
conditions that would allow such a separation. Since the 
longitudinal-photon interactions vanish in the limit $Q^2 \to 0$, the 
$Q^2$ dependence of interaction rates could offer an alternative probe.
Assuming that the resolved longitudinal interactions are gradually turned on 
up to $Q^2 \sim m_{\rho}^2$, however, over that range also the interactions 
of the transverse photons partly change character, and furthermore direct 
longitudinal photons begin to contribute. It is thus not clear to what 
extent the structure of the longitudinal photon can be probed separately, 
on top of everything else.

The recent presentation of a set of QCD-evolved parton distributions for 
the longitudinal photon has allowed a first assessment in this letter. 
(Other studies, without QCD evolution, were presented in \cite{ChyTas}.)
The $x$ dependence of the Ch\'yla PDF's clearly are different from those 
of the transverse photons. Therefore the ratio of the two cannot be modeled by
simple $x$-independent factors, the way we have tried in our previous articles.
It is then rather disappointing to note that most of these differences
are masked in typical experimental quantities. In some instances, the resolved
longitudinal-photon contribution itself is rather small relative to other
event categories. Even where it is not so small, in quantities like jet rates 
and total $\gast\gast$ cross sections, the smearing in $x$ and $\mu^2$ 
is significant. 

Thus it comes that a simple factor like our 
$r_1 = a 4 m_{\rho}^2 Q^2/(m_{\rho}^2 + Q^2)^2$, with $a \approx 0.5$,
gives quite a decent description of the resolved longitudinal effects  
when photon virtualities are of the order of $m_{\rho}^2$. At larger
virtualities it is dampened too fast however. 
(When considering total cross sections, including elastic, diffractive and
low-$\pT$ events, it was found to agree with data at large photon 
virtualities~\cite{ourtot}. A longitudinal PDF will not give any new 
insight here however, since no perturbative scale can be associated with 
the scattering process and, moreover, other descriptions are used.)
On the other hand, the $r_2 = a 4 Q^2/(m_{\rho}^2 + Q^2)$ factor was found to
give a decent description of the longitudinal effects in $\gast\gast$ events,
characterized by larger $Q^2$ than the jet studies, while the $r_1$ failed to 
do so. A hybrid of the two, e.g. $a 4 Q^2/(m_{\rho}^2 + b Q^2)$, could 
accommodate for the $r_1$ behaviour at $Q^2 \approx m_{\rho}^2$ 
($b \approx 2-3$) and would approach a constant value for large $Q^2$, 
similar to $r_2$. 
With $R=a 4 Q^2/(m_V^2 + b Q^2)$, $a=0.5$ and $b=2$,  
approximately the same result as $r_1$ in the $\gast\p$ jet cross 
section and as $r_2$ in $\e^+\e^- \ra \e^+\e^- + \mr{hadrons}$ ($\gast\gast$) 
are obtained (Fig.~\ref{fig:ee+gp}), i.e. in decent agreement with the parton 
distribution fraction $R_\mr{PDF}$ for both cases. 
It appears likely, but remains to be demonstrated, that a simple factor of 
this kind could work over a broad kinematical range, for various observables.

The other simple alternatives, $R_1$ and $R_2$, are
disfavoured. In particular, a significant $\mu^2$ dependence of 
$f_i^{\gast_{\mr{L}}}/f_i^{\gast_{\mr{T}}}$ is only present at large $x$,
and here direct processes may be expected to dominate the data samples.  

In summary, the bad news is that the experimental studies of 
$f_i^{\gast_{\mr{L}}}(x_i, \mu^2, Q^2)$ may turn out to be very difficult.
The good news is that the uncertainty from a non-understanding of the
resolved longitudinal photon now can be reduced, which should simplify 
the task of exploring other aspects of photon physics.

\subsection*{Acknowledgement}

We thank J. Ch\'yla for providing the parameterizations of his
distributions and for helpful correspondence.


\begin{thebibliography}{99}
%
%
%

\bibitem{ourjet}
C. Friberg and T. Sj\"ostrand,
\Journal{\EPJC}{13}{2000}{151}
(hep-ph/9907245).

\bibitem{ourtot}
C. Friberg and T. Sj\"ostrand,
LU TP 00--29 (hep-ph/0007314).

\bibitem{longgastpol}
CHIO Collaboration, W.D. Shambroom et al., \Journal{\PRD}{26}{1982}{1};\\
NMC Collaboration, P. Amaudruz et al., \Journal{\ZPC}{54}{1992}{239};\\
NMC Collaboration, M. Arneodo et al., \Journal{\NPB}{429}{1994}{503};\\
E665 Collaboration, M.R. Adams et al., \Journal{\ZPC}{74}{1997}{237};\\
ZEUS Collaboration, J. Breitweg et al., \Journal{\EPJC}{6}{1999}{603},\\
\Journal{\EPJC}{12}{2000}{393};\\
H1 Collaboration, C. Adloff et al., \Journal{\EPJC}{13}{2000}{371};\\
HERMES Collaboration, K. Ackerstaff et al., hep-ex/0002016.

\bibitem{longgastR}
BCDMS Collaboration, 
A.C. Benvenuti et al., \Journal{\PLB}{223}{1989}{485};\\
H1 Collaboration, C. Adloff et al., \Journal{\PLB}{393}{1997}{452};\\
HERMES Collaboration, K. Ackerstaff et al., 
\Journal{\PLB}{475}{2000}{386}.

\bibitem{ChylaFl}
J. Ch\'yla, PRA-HEP 00-03 (hep-ph/0006232).

\bibitem{siggaga}
V.M.~Budnev, I.F.~Ginzburg, G.V.~Meledin and V.G.~Serbo,\\ 
\Journal{\PRP}{15}{1974}{181};\\
V.N.~Baier, E.A.~Kuraev, V.S.~Fadin and V.A.~Khoze,\\ 
\Journal{\PRP}{78}{1981}{293}.

\bibitem{longgastQED}
A.S. Gorski, B.L. Ioffe, A.Yu. Kodjamirian and A. Oganesian,\\
\Journal{\ZPC}{44}{1989}{523}.

\bibitem{IngSch}
G. Ingelman and P.E. Schlein, \JournalPLB{152}{1985}{256}.

\bibitem{SaSmodel}
G.A. Schuler and T. Sj\"ostrand, 
\Journal{\PLB}{300}{1993}{169}, \\
\Journal{\NPB}{407}{1993}{539},
\Journal{\ZPC}{73}{1997}677.

\bibitem{pythia} 
T. Sj\"ostrand, \Journal{\CPC}{82}{1994}{74};\\
T. Sj\"ostrand et al., in preparation;\\
http://www.thep.lu.se/$\sim\,$torbjorn/Pythia.html.

\bibitem{SaSpdf}
G.A. Schuler and T. Sj\"ostrand, 
\Journal{\ZPC}{68}{1995}{607}, \\
\Journal{\PLB}{376}{1996}{193}.

\bibitem{DGLAP}
V.N. Gribov and L.N. Lipatov, \Journal{\SJNP}{15}{1972}{438 and 675};\\
G.~Altarelli and G.~Parisi, \Journal{\NPB}{126}{1977}{298};\\
Yu.L.~Dokshitzer, \Journal{\SPJP}{46}{1977}{641}.

\bibitem{sigmaTL}
L.N.~Hand, \Journal{\PRV}{129}{1963}{1834}.

\bibitem{siggap}
G. Altarelli and G. Martinelli, \JournalPLB{76}{1978}{89};\\
A. Mend\'ez, \Journal{\NPB}{145}{1978}{199};\\
R. Peccei and R. R\"uckl, \Journal{\NPB}{162}{1980}{125};\\
Ch. Rumpf, G. Kramer and J. Willrodt, \Journal{\ZPC}{7}{1981}{337}.

\bibitem{RinDISthy}
S.~R.~Mishra and F.~Sciulli,
\Journal{\PLB}{244}{1990}{341}.

\bibitem{xgamma}
ZEUS Collaboration, M. Derrick et al., \Journal{\PLB}{322}{1994}{287};\\
H1 Collaboration, T. Ahmed et al., \Journal{\NPB}{445}{1995}{195}.

\bibitem{CTEQ5L}
CTEQ Collaboration, H.~L.~Lai et al.,  
\Journal{\EPJC}{12}{2000}{375}.

\bibitem{L3}
L3 Collaboration, M.~Acciarri et al.,
\Journal{\PLB}{453}{1999}{333}.

\bibitem{ChyTas}
J. Ch\'yla and M. Tasevsk\'y, PRA-HEP 99-07 (hep-ph/9912514),\\
PRA-HEP 00-01 (hep-ph/0003300).

\end{thebibliography}
\end{document}